\documentclass[12pt]{iopart}

\newcommand{\be}{\begin{equation}}
\newcommand{\ee}{\end{equation}}
\newcommand{\bea}{\begin{eqnarray}}
\newcommand{\eea}{\end{eqnarray}}
\newcommand{\beaa}{\begin{eqnarray*}}
\newcommand{\eeaa}{\end{eqnarray*}}
\newcommand{\nn}{\nonumber \\}

\begin{document}

\tolerance=5000

\title{Dark energy with time-dependent equation of state}

\author{O.G. Gorbunova}
\address{Lab. for Fundamental Studies,
Tomsk State Pedagogical University, 634041 Tomsk, Russia}

\author{A.V. Timoshkin}
\address{Tomsk State Pedagogical University, 634041 Tomsk, Russia}



\begin{abstract}

The accelerating Friedmann flat universe, filled with an ideal fluid
with a linear (oscillating) inhomogeneous equation of state (EoS)
depending on time, is reviewed. The equations of motions are solved.
It is shown that in some cases there appears a quasi-periodic
universe, which repeats the cycles of phantom-type space
acceleration \footnote{This article is {\bf dedicated to 70th
aniversary of Professor Iver Brevik}}.

\end{abstract}

\maketitle

\section{Introduction}

As is known, the present universe is subject to an acceleration,
which can be explained in terms of an ideal fluid(dark energy)
weakly interacting with usual matter and which has an uncommon
equation of state. The pressure of such an ideal dark energy fluid
is negative. In the present work we study a model where there is an
ideal fluid with an inhomogeneous equation of state,
$p=w(t)\rho+\Lambda(t)$ in which the parameters $w(t)$ and
$\Lambda(t)$ depend linearly on time. In another version, these
parameters are oscillating in time. Ideal fluids with an
inhomogeneous equation of state were introduced in \cite{1} (see
also examples discussed in \cite{2}). We show that, depending on the
choices for the input parameters, it is possible for the universe to
pass from the phantom era to the non-phantom era, implying the
appearance of a cosmological singularity. Also possible are cases of
quasi-periodic changes of the energy density and of the Hubble
constant, with the appearance of quasi-periodic singularities, and
the appearance of cosmological singularities.

The particular kind of equation of state in the present paper is one
alternative amongst a variety of possibilities, proposed recently to
cope with the general dark energy problem. Different examples
include imperfect equation of state \cite{4}, general equation of
state \cite{5}, inhomogeneous equation of state \cite{1}, \cite{6}
including time-dependent viscosity as a special case \cite{7}, and
multiple-Lambda cosmology \cite{8}. In the next section, a linear
inhomogeneous EoS ideal fluid in a FRW universe is studied. In
section 3, an oscillating inhomogeneous EoS ideal fluid is
investigated. Some discussion is presented in the last section.

\section{Inhomogeneous equation of state for the universe and its solution}
Let us assume that the universe is filled with an ideal fluid (dark
energy) obeying an inhomogeneous equation of state (see also
\cite{9}),
 \be \label{1} p=w(t)\rho+\Lambda(t) , \ee

where $w(t)$ and $\Lambda(t)$ depend on the time t. This equation of
state, when $\Lambda(t)=0$ but with $\omega(t)$ a function of time,
examined in \cite{9} and \cite{10}.

Let us write down the law of energy conservation:

\be \label{2} dot{\rho}+3H(p+\rho)=0 , \ee

and Friedman's equation:

\be \label{3} \frac{3}{\chi^{2}}H^{2}=\rho , \ee

where $\rho$ is the energy density, $p$- the pressure,
$H=\frac{\dot{a}}{a}$ the Hubble parameter, $a(t)$- the scale factor
of the three-dimensional flat Friedman universe, and $\chi$ - the
gravitation constant.

Taking into account equations (\ref{1}), (\ref{2}) and (\ref{3}), we
obtain the gravitational equation of motion :

\begin{equation}
\dot{\rho}+\sqrt{3}\kappa(1+\omega(t))\rho^{3/2}+\sqrt{3}\kappa
\rho^{1/2}\Lambda(t)=0.\label{4}
\end{equation}

Let us suppose in the following that both functions $w(t)$ and
$\Lambda(t)$ linearly on time:

\begin{equation}
\omega(t)=a_{1}t+b, \label{5}
\end{equation}

\begin{equation}
\Lambda(t)=ct+d.\label{6}
\end{equation}

This kind of behaviour may be the consequence of a modification of
gravity (see \cite{11} for a review).

Let $\Lambda(t)=0$, $\omega(t)=a_{1}t+b$. In this case the energy
density takes the form:

\begin{equation}
\rho(t)=\frac{4(2a_{1}+1)^{2}}{3\chi^{2}}\cdot\frac{(a_{1}t+b+1)^{2/a_{1}}}{[(a_{1}+b+1)^{\frac{1}{a_{1}}+2}+S]^{2}},
\label{7}
\end{equation}

Hubble's parameter becomes:

\begin{equation}
H(t)=\frac{2}{3}(2a_{1}+1)\cdot\frac{(a_{1}t+b+1)^{1/a_{1}}}{(a_{1}+b+1)^{\frac{1}{a_{1}}+2}+S},
\label{8}
\end{equation}
where $S$ is an integration constant.

The scale factor takes the following form:

$a(t)=\exp^{\frac{2}{3}(2a_{1}+1)I}$,

where

\begin{eqnarray}
&&I=\frac{(-1)^{a_{1}}}{(2a_{1}+1)S^{a_{1}}}\ln
|(a_{1}t+b+1)^{\frac{1}{a_{1}}+2}+S|-\frac{1}{(2a_{1}+1)S^{a_{1}}}\cdot\\\nn
&& \sum^{a_{1}-1}_{k=0}
\cos\frac{(a_{1}+1)(2k+1)}{2a_{1}+1}\cdot\pi\cdot\ln([(a_{1}t+b+1)^{\frac{2}{a_{1}}}+S]^{2}-\\\nn
&&2S(a_{1}t+b+1)^{\frac{1}{a_{1}}}\cdot\cos\frac{2k+1}{2a_{1}+1}\pi+S^{2})+\frac{2}{(2a_{1}+1)S^{a}}\cdot\\\nn
&&\sum^{a_{1}-1}_{k=0}sin\frac{(a_{1}+1)(2k+1)}{2a_{1}+1}\cdot\pi\cdot
\arctan\frac{(a_{1}t+b+1)^{\frac{1}{a_{1}}}-S\cdot\cos\frac{2k+1}{2a_{1}+1}\cdot\pi}
{S\cdot\sin\frac{2k+1}{2a_{1}+1}\cdot\pi},\label{10} \end{eqnarray}

$1\leq a_{1} \leq 2a_{1}-1.$

At $t_{1}=-\frac{b+1}{a_{1}}$ or
$t_{2}=\frac{1}{a_{1}}(\frac{Sa_{1}}{a_{1}+1})^{\frac{1}{2+\frac{1}{a_{1}}}}
-\frac{b+1}{a_{1}}$, one has $\dot{H}=0$. With $a_{1}>0$, $b>-1$ and
$t<t_{2}$, one gets $\dot{H}>0$ , that is, the accelerating universe
is in the phantom phase (see, for example, \cite{12}), and with
$t>t_{2}$, one gets $\dot{H}<0$ , the universe is in the non-phantom
phase. At the moment when the universe passes from the phantom to
the non-phantom era, Hubble's parameter equals

\begin{equation}
H_{m}=\frac{2}{3S}\sqrt[2a_{1}+1]{a_{1}(a_{1}+1)^{2a_{1}}}.
\label{11}
\end{equation}

In the phantom phase $\dot{\rho}>0$ the energy density grows; in the
non-phantom phase $\dot{\rho}<0$ the energy density decreases.
However, the derivative of the scale factor $\dot{a}>0$, therefore
the universe expands. Note, as has been shown in \cite{9}, that in
the phantom phase the entropy may become negative. If
$t\rightarrow+\infty$, then $H(t)\rightarrow 0$ and
$\rho(t)\rightarrow 0$, so that the phantom energy decreases.

\section{Inhomogeneous oscillating equation of state}

Instead of assuming the form Eq. (\ref{5}) and Eq. (\ref{6})for the
time-dependent parameters, one might assume that there is an
oscillating dependence on time. Let us investigate the following
form: $w(t)=-1+\omega_{0}cos\omega t$. From equations (\ref{3}) and
(\ref{4}) we get

\begin{equation}
H=\frac{2\omega}{3(\omega_{1}+\omega_{0}sin\omega t)}, \label{12}
\end{equation}

where $\omega_{1}$ is an integration constant. If
$|\omega_{1}|<\omega_{0}$, the denominator can in this case be zero,
what implies a future cosmological singularity. But if
$|\omega_{1}|>\omega_{0}$, singularity is absent.

In view of the fact that (cf. \cite{9})

\begin{equation}
\dot{H}=-\frac{2\omega^{2}\omega_{0}cos\omega
t}{3(\omega_{1}+\omega_{0}sin\omega t)}, \label{13}
\end{equation}

we see that with $\omega_{0}cos\omega t<0$ ($\omega_{0}cos\omega
t>0$)the universe is located in the phantom (non-phantom) phase,
corresponding respectively to $\dot{H}>0$ ($\dot{H}<0$). If the
oscillation period of the universe is large, it is possible to have
a unification of inflation and phantom dark energy \cite{12,13}. The
density of dark energy takes the form

\begin{equation}
\rho(t)=\frac{4\omega^{2}}{3\chi^{2}(\omega_{1}+\omega_{0}sin\omega
t)^{2}}, \label{14}
\end{equation}
this being a periodic function so that the universe oscillates
between the phantom and non-phantom eras.

Let us assume now that $\Lambda(t)\neq0$. For simplicity we take
$\Lambda(t)=\Lambda_{0}sin\omega t$, i.e. a periodic function. If
$\Lambda_{0}(t)<0$, equation (\ref{4}) has the following solution:

\begin{equation}
\frac{\sqrt{\rho(t)}+\sqrt{\frac{|\Lambda_{0}|}{\omega_{0}}}}{\sqrt{\rho(t)}-\sqrt{\frac{|\Lambda_{0}|}{\omega_{0}}}}=
exp[\sqrt{3\chi^{2}|\Lambda_{0}|\omega_{0}}(\frac{sin\omega
t}{\omega}+C_{1})], \label{15}
\end{equation}

where $C_{1}$ is an integration constant.

 Finally, we obtain for
the energy density:

\begin{equation}
\rho(t)=
\{\frac{\sqrt{\frac{|\Lambda_{0}|}{\omega_{0}}}exp[\sqrt{3\chi^{2}|\Lambda_{0}|\omega_{0}}(\frac{sin\omega
t}{\omega}+C_{1})]+\sqrt{\frac{|\Lambda_{0}|}{\omega_{0}}}}
{exp[\sqrt{3\chi^{2}|\Lambda_{0}|\omega_{0}}(\frac{sin\omega
t}{\omega}+C_{1})]-1}\}. \label{16}
\end{equation}

The Hubble parameter becomes, according to (\ref{3}),

\begin{equation}
H(t)=\sqrt{\frac{\chi^{2}\rho(t)}{3}}. \label{17}
\end{equation}

At the moments when the denominator of (\ref{16}) is zero, the
energy density diverges. This corresponds to a future cosmological
singularity. Depending on the choice of parameters in the equation
of state for the dark energy, $H(t)$ can thus correspond to either a
phantom, or a non-phantom, universe. In both cases the universe
expands with (quintessential or super) acceleration.

\section{Summary}
In this work we have studied a model of the universe in which there
is a linear inhomogeneous equation of state, with a linear or
oscillating dependence on time. The consequences of various choices
of parameters in the linear functions are examined: there may occur
a passage from the non-phantom era of the universe to the phantom
era, resulting in an expansion and a possible appearance of
singularities. In the absence of inhomogeneous terms it is possible
to have a repetition of the passage process. When the universe goes
from the phantom era to the non-phantom era with expansion one may
avoid singularities, or there may appear singularities, but the
passage occurs without repetition. The presence of a linear
inhomogeneous term in the equation of state leads either to a
compression of the universe in the evolution process or to a
quasi-periodic change in the energy density and in the Hubble
parameter, and also to a quasi-periodic appearance of singularities.
By this the universe either passes into the non-phantom era, or
stays within the same era as it was originally. Thus, the universe
filled with an inhomogeneous time-dependent equation-of-state ideal
fluid may currently be in the acceleration epoch of quintessence or
phantom type. Moreover, it is easy to see that the effective value
of the equation-of-state parameter may easily be adjusted so as to
be approximately equal to -1 at present, what corresponds to current
observational bounds.

\section{Acknowledgment}
We thank Sergei Odintsov for very valuable information.

This article represents brief review of our paper \cite{14}.

This work is supported by Grant the Scientific School LRSS PROJECT
N.2553.2008.2., and also by Grant RFBR 06-01-00609 .

\end{document}